\documentclass[a4paper,fleqn,usenatbib]{mnras}

\usepackage{newtxtext,newtxmath}
\usepackage{txfonts}

\usepackage[T1]{fontenc}
\usepackage{ae,aecompl}

\usepackage{graphicx}	
\usepackage{pstricks}
\usepackage{color}
\usepackage{ulem}
\usepackage{enumitem}

\graphicspath{{figures/}}

\title[Transits of solar system objects in Kepler images]{Prediction of transits of solar system objects in \textit{Kepler}/K2 images: An extension of the Virtual Observatory service SkyBoT}

\author[J. Berthier et al.]{
	J. Berthier,$^{1}$\thanks{E-mail: berthier@imcce.fr}
  	B. Carry,$^{1,2}$
  	F. Vachier,$^{1}$
  	S. Eggl,$^{1}$
  	and A. Santerne,$^{3}$
	\\
	$^{1}$IMCCE, Observatoire de Paris, PSL Research University, CNRS, Sorbonne Universit\'es, UPMC Univ Paris 06, Univ Lille, France\\
	$^{2}$Laboratoire Lagrange, Universit\'e de Nice-Sophia Antipolis, CNRS, Observatoire de la C\^ote d'Azur, France\\
	$^{3}$Instituto de Astrof\'isica e Ci\^encias do Espa\c co, Universidade do Porto, CAUP, Rua das Estrelas, 4150-762 Porto, Portugal\\
}

\date{Accepted XXX. Received YYY; in original form ZZZ}

\pubyear{2016}

\begin{document}
	\label{firstpage}
	\pagerange{\pageref{firstpage}--\pageref{lastpage}}
	\maketitle
	
	\begin{abstract}
      All the fields of the extended space mission \textit{Kepler}/K2 are located
      within the ecliptic. Many solar system objects thus cross the K2 stellar masks 
      on a regular basis. We aim at providing to the entire community a simple tool 
      to search and identify solar system objects serendipitously observed by \textit{Kepler}.
      The SkyBoT service hosted at IMCCE provides a Virtual Observatory (VO) compliant 
      cone-search that lists all solar system objects present within a field of view 
      at a given epoch. To generate such a list in a timely manner, ephemerides are 
      pre-computed, updated weekly, and stored in a relational database to ensure a 
      fast access. The SkyBoT Web service can now be used with \textit{Kepler}. Solar 
      system objects within a small (few arcminutes) field of view are identified and 
      listed in less than 10\,s. Generating object data for the entire K2 field of view 
      (14\degr) takes about a minute. This extension of the SkyBot service opens new 
      possibilities with respect to mining K2 data for solar system science, as well 
      as removing solar system objects from stellar photometric time-series.
	\end{abstract}
	


\section{Introduction}

  \indent The NASA Discovery mission \textit{Kepler} was launched in 2009, with 
  the aim of detecting exoplanets from the photometric signature of their transit 
  in front of their host star \citep{2009Sci...325..709B}. Following the second 
  failure of a reaction wheel in May 2013, the original field of view (FoV) in 
  Cygnus could not be fine pointed anymore. An extension of the mission, dubbed 
  K2 \citep{2014PASP..126..398H}, was designed to be a succession of 3-month 
  long campaigns, where the spacecraft's FoV scans the ecliptic plane. 
  This mode of operations implies that many solar system objects (SSOs) cross 
  the subframes centered on K2 mission targets. Following a visual inspection 
  of the K2 engineering FoV, \citet{2015-AJ-149-Szabo} reported that SSOs had 
  crossed half of the 300 stars monitored over the 9 days of engineering 
  observations. \\

  \indent Owing to the large number of stellar targets in each K2 campaign, the 
  likelihood of observing SSOs at any single epoch is indeed high. Given a typical 
  mask size around each target of 15x15 pixels or 1x1 arcmin for between 10,000 and 
  30,000 stellar targets, the filling factor\footnote{The fraction of the K2 FoV 
  that is actually downlinked.} of K2 entire FoV ranges from 3\% to 10\% (Table~\ref{tab:num}). 
  A corresponding fraction of the SSOs that cross K2 FoVs are within a target mask 
  at each instant, from a few tens of minutes for a near-Earth object to 
  approximately 6\,h for a main-belt asteroid, and up to several days for a 
  Trojan or a transneptunian object. Over a whole campaign, the cumulative probability 
  to observe these SSOs get close to one, as the different target masks, stacked 
  over ecliptic longitude, almost fill entirely the range of ecliptic latitudes 
  within K2 field of views (Table~\ref{tab:num}). Each SSO has thus only a few 
  percent chance to dodge all the target masks as it crosses K2 field of view 
  (Table~\ref{tab:num}). Several programs dedicated to planetary science have been
  already carried out by K2, like characterization of the rotation period of transneptunian 
  objects \citep{2015-ApJ-Pal}. The giant planet Neptune and its satellites
  were also observed in C3, and Uranus will be in C8.\\

\begin{figure}
	\includegraphics[width=\columnwidth]{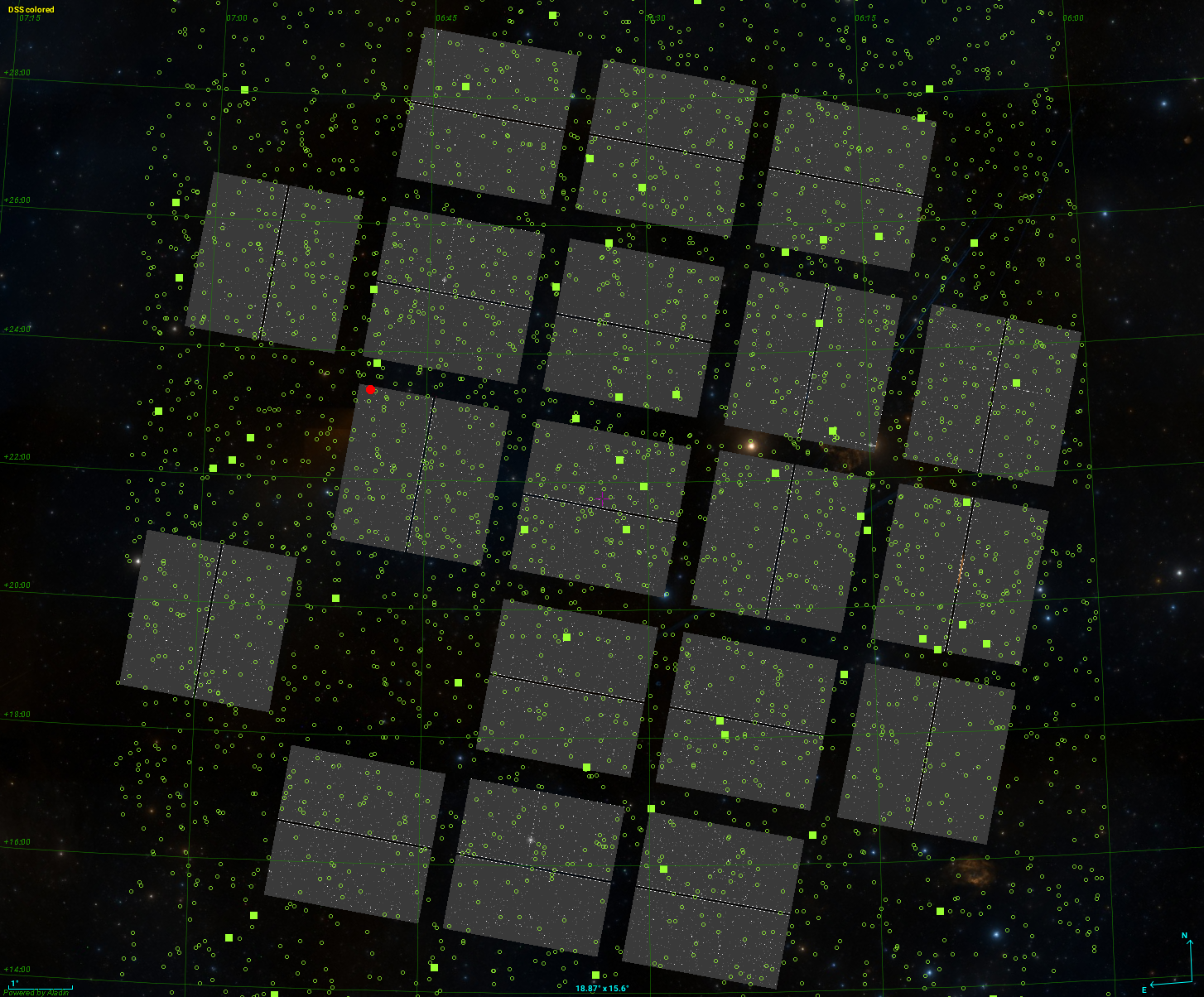}
	\caption{K2 full frame image taken on 2014, March, the 11\textsuperscript{th}, 
		at 23:27:23.77 UTC (mid-exposure), over-plotted on the DSS colored view, 
		displayed by Aladin. All the 3136 known SSOs brighter than V\,$\leq$\,20 
		(among 9702) present within the FoV reported by SkyBoT are represented, by the 
		green circles for asteroids (and solid squares for V\,$\leq$\,16.5 ), and by 
		the red dot for a comet (84P, V\,=\,18.8).}
	\label{fig:k2-C0}
\end{figure}

  \indent Considering the typical magnitude of K2 stellar targets (80\% of the stars 
  have a V\,$\leq$\,15-16), and the typical K2 photometric precision of a few hundreds
  ppm, many SSOs will be imaged together with the stars. At any instant several thousands
  of SSOs with V\,$\leq$\,20 lay within K2 entire field of view
  (e.g., Fig.~\ref{fig:k2-C0}).
  A magnitude 20 asteroid will contribute to the star signal at a level of 1000 ppm, 
  and is, therefore, easily detectable. \\

  \indent There is a twofold interest in having a simple tool to predict 
  \textsl{encounters} between stars and SSOs: 
  \begin{itemize}[topsep=0pt,itemsep=-1ex,partopsep=1ex,parsep=1ex]
    \item[$\circ$] The K2 community profits from identifying any encounters that 
      add undesirable signals, hence photon noise, to stellar light curves, at 
      non-negligible levels.
    \item[$\circ$] The solar system community profits, as each encounter provides 
      a short light curve (typical a couple of hours) of an SSO with excellent 
      photometric accuracy. On average, ten encounters per campaign can be 
      expected (Table~\ref{tab:num}).\\
  \end{itemize}
  
  \indent To cater to those demands, we present an extension of our Virtual 
  Observatory (VO) tool SkyBoT \citep{2006ASPC..351..367B-Berthier}, hosted at 
  IMCCE. This tool is web based, open-access, and provides a simple way to identify 
  all the SSOs present within a field of view at a given epoch. This article 
  is organized as following: in Section~\ref{sec:skybot} we describe the SkyBoT 
  service, its algorithm and access, and we show a pair of examples in 
  Section~\ref{sec:examples}. 

\begin{table}
  \centering
  \caption{Number of K2 stellar targets, fraction of the total field of view 
  	downlinked to Earth, filling fraction of ecliptic latitudes ($\beta_f$),
   expected average number and standard deviation of stellar encounters for
   each SSO ($\mu_e$ and $\sigma_e$), for each campaign (up to C7).}
  \label{tab:num}
  \begin{tabular}{lrccrr}
    \hline
    Campaign  & Targets & Area (\%) & $\beta_f$  (\%) & $\mu_e$ & $\sigma_e$ \\
    \hline
    C0 &  7756 &   2.90 &  94.16 &  4.3 &  2.7 \\ 
    C1 & 21647 &   8.09 &  98.25 & 11.8 &  5.4 \\ 
    C2 & 13401 &   5.01 &  96.53 &  7.4 &  4.4 \\ 
    C3 & 16375 &   6.12 &  97.94 &  9.1 &  4.8 \\ 
    C4 & 15781 &   5.90 &  98.18 &  8.7 &  4.2 \\ 
    C5 & 25137 &   9.40 &  98.68 & 13.8 &  6.3 \\ 
    C6 & 27289 &  10.20 &  98.91 & 14.9 &  6.2 \\ 
    C7 & 13261 &   4.96 &  96.74 &  7.3 &  4.9 \\ 
    \hline
  \end{tabular}
\end{table}


\section{SkyBoT: The VO Sky Body Tracker}
\label{sec:skybot}

  \indent The typical queries to astronomical catalogs are so-called 
  \textsl{cone searches}, in which all targets within a given field of view 
  are returned. This is mostly adapted to objects with fixed coordinates,
  such as stars and galaxies, their parallax and proper motion being much 
  smaller than the field of view. But the coordinates of objects in our 
  solar system constantly change and cone searches cannot use pre-defined
  catalogs. As a result, most tools for source identification fail to
  associate the observed SSO with a known source. The SkyBoT service provides 
  a solution by pre-computing ephemerides of all the known SSOs, and storing 
  them in a relational database for rapid access upon request.

  \subsection{Ephemerides computation and SkyBoT algorithm\label{ssec:calc}}

     \indent Among other services, the Institut de m\'ecanique c\'eleste et de 
     calcul des \'eph\'em\'erides (IMCCE) produces the French national ephemerides 
     under the supervision of the Bureau des longitudes. The development and 
     maintenance of ephemerides tools for the astronomical community is also a 
     part of its duties. As such, the institute offers online computation of solar 
     system object ephemerides through a set of Web  services\footnote{\href{http://vo.imcce.fr}{http://vo.imcce.fr/webservices/}}.\\

    \indent The ephemerides of planets and small solar system objects are 
    computed in the ICRF quasi-inertial reference frame taking into account 
    perturbations of the 8 planets, and post-Newtonian corrections. The geometric 
    positions of the major planets and the Moon are provided by INPOP planetary 
    theory \citep{2014-SciNote-Fienga}. Those of small SSOs (asteroids, comets, 
    Centaurs, trans-neptunian objects) are calculated by numerical integration of 
    the N-body perturbed problem \citep[Gragg-Bulirsch-Stoer algorithm, see][]{Bulirsch66, Bulirsch80}, 
    using the latest published osculating elements, from the \texttt{astorb} 
    \citep{1993LPICo.810...44B-Bowell} and \texttt{cometpro} \citep{1996IAUS..172..357R-Rocher} 
    databases. The overall accuracy of asteroid and comet ephemerides provided by
    our services are at the level of tens of milli-arcseconds, mainly depending on 
    the accuracy of the minor planet's osculating elements. The positions of natural 
    satellites are obtained thanks to dedicated solutions of their motion, e.g. 
    \citet{2004-AA-420-Lainey, 2004-AA-427-Lainey, 2007-AA-465-Lainey} for Mars 
    and Jupiter, \citet{1995-AA-297-Vienne} for Saturn, \citet{1987-AA-188-Laskar} 
    for Uranus, and \cite{LeGuyader93} for Neptune's satellites. \\

    \indent The ephemerides of all the known objects of our solar System are 
    recomputed on a weekly basis, for a period which extends from the end of the
    19\textsuperscript{th} century (1889-11-13) to the first half of the 
    21\textsuperscript{st} century (2060-03-21), and stored with a time step 
    of 10 days in a hierarchical tree structure supported by nodes based on geocentric
    equatorial coordinates. For each cone search, this database is queried, and 
    all the targets expected to be within the field of view are listed. Their 
    topocentric ephemerides for the exact requested time are then computed on the fly.  \\

    \indent The apparent topocentric celestial coordinates (i.e. relative to the true
    equator and equinox of the date) are computed by applying light aberration,
    precession, and nutation corrections to the observer-target vector. The 
    coordinates of the topocenter can either be provided directly by users (longitude, 
    latitude, altitude), or by using the observatory code provided by IAU Minor Planet
    Center\footnote{\href{http://www.minorplanetcenter.net/iau/lists/ObsCodesF.html}{http://www.minorplanetcenter.net/iau/lists/ObsCodesF.html}}
    for listed observatories. \\

    \indent The SkyBoT service was released in 2006 \citep[][]{2006ASPC..351..367B-Berthier}. 
    It is mostly used to identify moving objects in images 
    \citep[e.g.][]{2009-EMP-105-Conrad, 2011ASPC..442..111D-Delgado, 2012-AA-544-Carry, 2013-AA-554-Bouy}, 
    and data mining of public archives 
    \citep[e.g.][]{2009AN....330..698V-Vaduvescu, 2011-AN-332-Vaduvescu, 2013-AN-334-Vaduvescu, 2016-Icarus--Carry}. 
    It responds to about 80,000 requests every month (more than 18 millions in 7 years),
    and has a typical response time of less than 10\,s for 95\% of requests. 


  \subsection{An extension to non Earth-bound geometries\label{ssec:exten}}

    \indent Owing to the large number of known SSOs (currently 700,000), and the 
    extended period of time that needs to be covered (from the first photographic 
    plates to the present), pre-computations are the key to a timely service.
    As the database of pre-computed ephemerides was ordered in a tree based on 
    equatorial coordinates (RA/Dec) to allow quick identification of potential 
    targets within a field of view, the service was limited to a single geometry. 
    The large parallax presented by objects within the solar system indeed implies 
    different equatorial coordinates depending on the position of the observer.
    The first releases of SkyBoT were thus limited to Earth geocenter, topocenters, 
    and low-orbit satellites such as the Hubble Space Telescope or the International 
    Space Station. \\

\begin{figure}
		\includegraphics[width=\columnwidth]{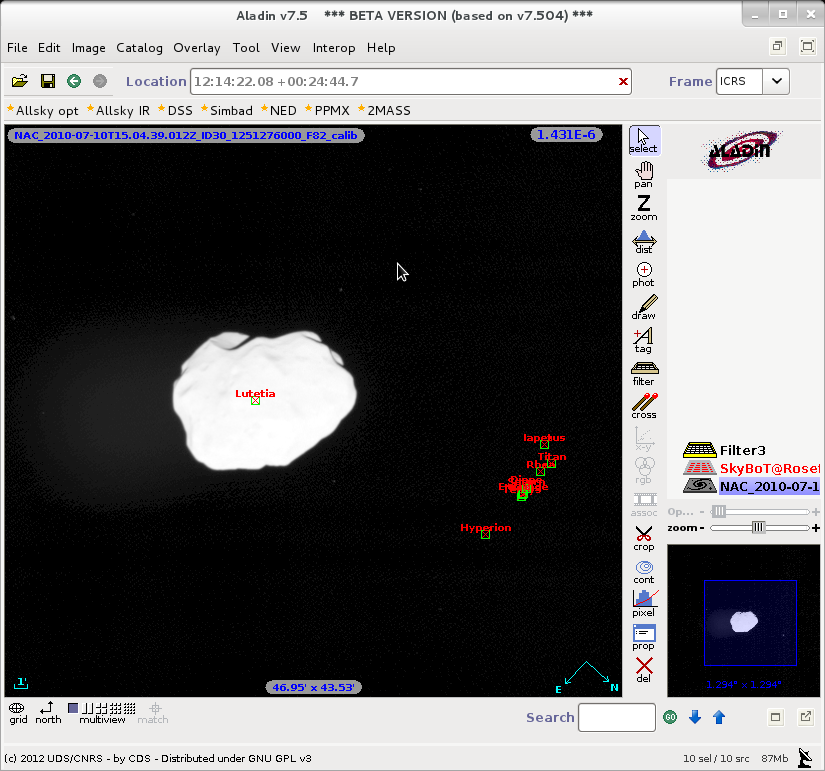}
		\caption{OSIRIS NAC image taken during the flyby of asteroid (21) Lutetia 
			by ESA \textit{Rosetta} space mission, on 2010, July, the 10\textsuperscript{th}, 
			at 15:04:30 UTC \citep{2011-Science-334-Sierks}, displayed in Aladin.
			A SkyBoT cone-search query correctly lists Lutetia, together with Saturn 
			and its satellites imaged in the background. Considering their dramatic 
			difference of distance to \textit{Rosetta} (36,000 km\,and 6.8\,au respectively),
			this example validates the SkyBoT upgrade to space missions.}
		\label{fig:rosetta}
	\end{figure}

    \indent In 2010, we started a new phase of the SkyBoT development to allow the 
    use of its cone-search method from other geometries. This was motivated by 
    availability of wide-field (2\degr$\times$2\degr~and 10\degr$\times$10\degr) 
    images taken by the OSIRIS camera on-board the ESA \textit{Rosetta} mission, 
    which is on an interplanetary trajectory crossing the asteroid main-belt, 
    between Mars and Jupiter. The great distance between the probe and the Earth, 
    combined with the proximity of SSOs implied observing geometries so different 
    that the Earth-bound database could not be used to search for and identify 
    targets correctly. This challenge was recently solved. An example validating 
    the corresponding update of the SkyBoT service is presented in Fig.~\ref{fig:rosetta}.\\

%
    \indent To preserve the fast response time of the service, a switch was set 
    in place, to redirect queries to different databases, one for each space probe. 
    These databases have smaller time coverage, corresponding only to the mission 
    lifetimes. The weekly computation of ephemerides is, therefore, not as CPU 
    intensive as for the main (Earth) database. There are currently two space 
    probes available: \textit{Rosetta} and \textit{Kepler}. The architecture of 
    SkyBoT after the update is such that we can add more space probes upon request: 
    any space mission located on a Earth leading or trailing orbit (e.g. Herschel), 
    or at L2 point (e.g. JWST, Euclid), or on a interplanetary trajectory
    (e.g. Cassini, JUNO) could be added, if desired by the community.

  \subsection{Access to the service\label{ssec:access}}

    \indent There are several ways to use the SkyBoT Web service. Users who may 
    want to discover the service can use a simple query form on the IMCCE's VO SSO
    portal\footnote{\href{http://vo.imcce.fr}{http://vo.imcce.fr}} or the 
    well-established Aladin Sky Atlas \citep{2000-AA-143-Bonnarel}. The service 
    is also fully compliant with VO standards, and thus, can be scripted in two 
    different ways: a) by writing a client to send requests to the SkyBoT server 
    and to analyze the response, or b) by using a command-line interface and a 
    data transfer program such as \texttt{curl} or \texttt{wget}.\\

    \indent 
    In all cases, three parameters must be passed to SkyBoT: the pointing 
    direction (RA/Dec), the epoch of observation, and the size of the field 
    of view. The typical response time for request from K2 point of view are
    of a few seconds for small field of view (target mask), and of about
    1\,min for the entire field of view of \textit{Kepler} of about 14\degr.

\section{Some examples}
\label{sec:examples}

	\begin{figure*}
		\includegraphics[width=\textwidth]{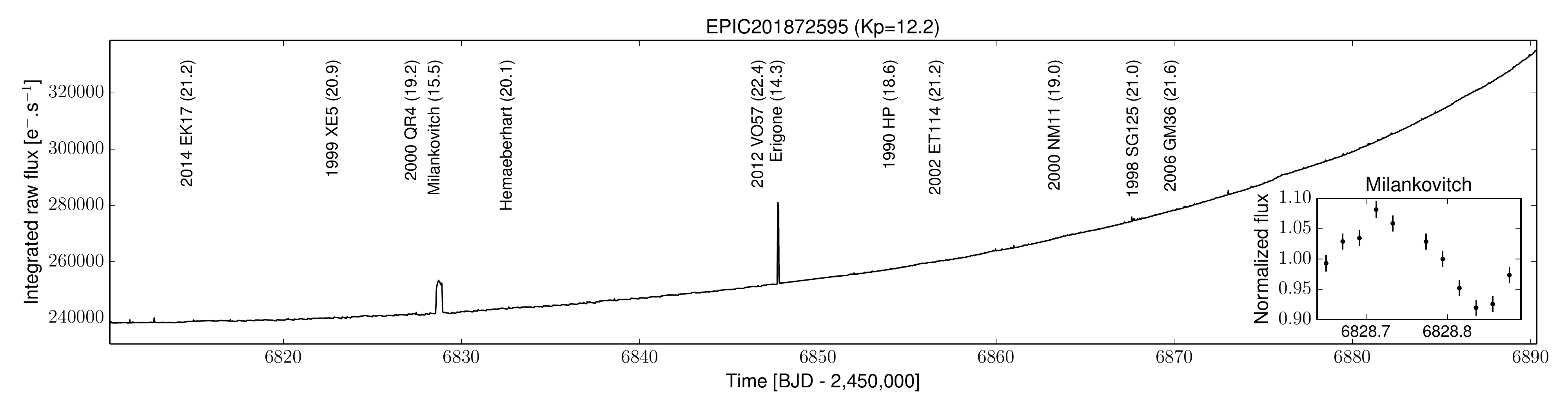}
		\caption{K2 raw light curve integrated over all pixels of the target EPIC 201872595
			(K$_{p}$ = 12.2) observed during Campaign \#1. The increase in flux along the 
			campaign is a systematic effect. The predicted transits of known SSOs down to 
			magnitude 22.5 are indicated together with their expected V magnitude. The 
			transit of two relatively bright SSOs, (1605) Milankovitch and (163) Erigone, 
			are clearly visible. The fainter SSOs also imprit a significant increase in 
			the observed flux as they pass into the target imagette. The inset in the bottom 
			right is a zoom on the transit of (1605) Milankovitch. It displays the 
			target-corrected and normalized flux of the SSO, and highlights the phase 
			rotation of the SSO.}
		\label{fig:star}
	\end{figure*}

	\begin{figure*}
		\includegraphics[width=\textwidth]{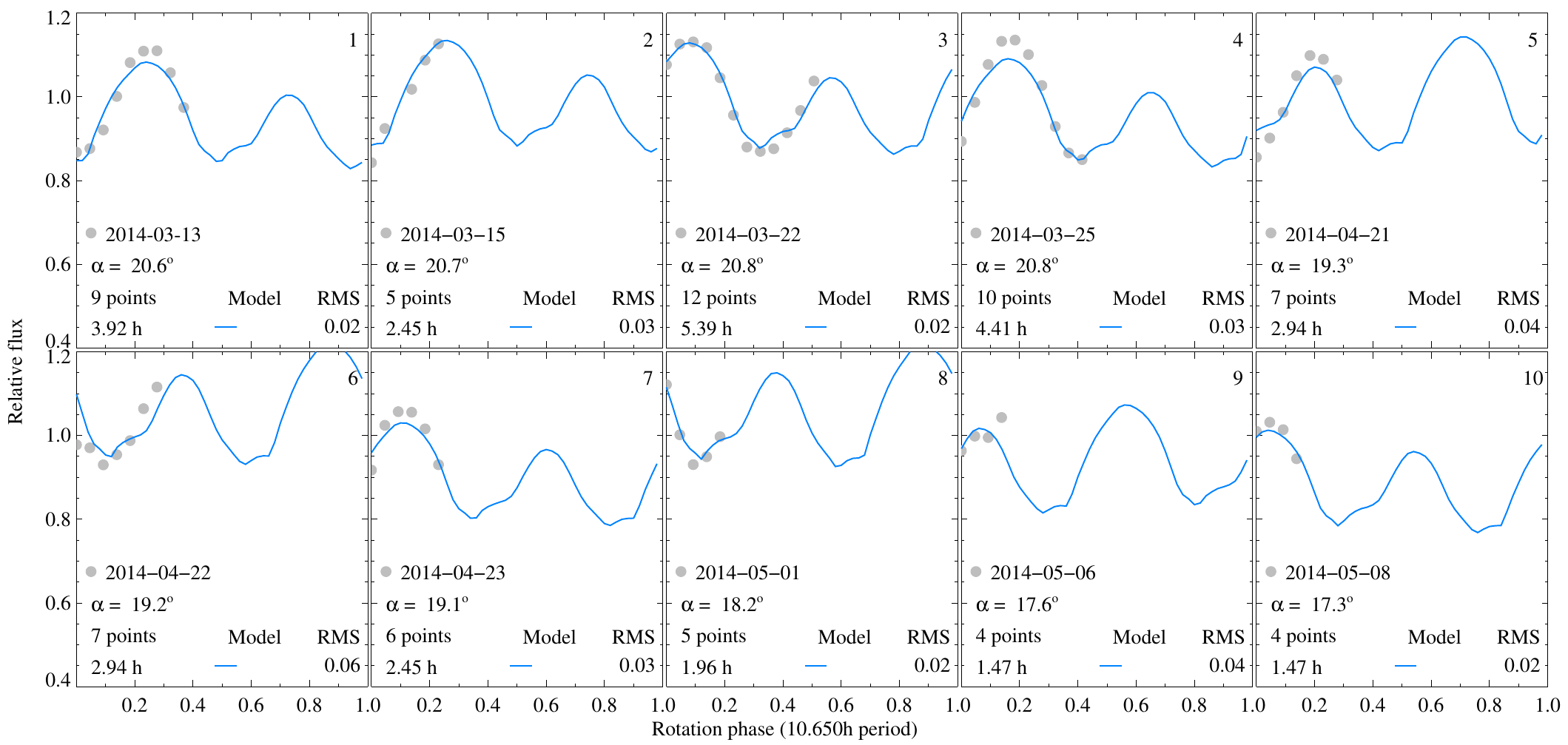}
		\caption{Example of asteroid light curves retrieved from K2 images. The grey 
			dots represent the measured photometry of (484) Pittsburghia, and the blue
			curves stand for the synthetic light curves obtained from the 3-D shape model 
			of the asteroid by \citet{2009-AA-493-Durech} and \citet{2011-AA-530-Hanus}.
			The residuals between observed and modeled points are of 0.03 magnitude on 
			average, as reported on each graph.}
		\label{fig:sso}
	\end{figure*}

  \indent We now present a couple of examples of the typical usage of the SkyBoT 
  service for K2. In Fig.~\ref{fig:k2-C0}, we show a full frame image from C0, 
  together with the result of a SkyBoT request: among the 9702 SSOs located
  in the FoV at that time, 3136 are brighter than V\,$\leq$\,20, and about 50 are 
  brighter than V\,$\leq$\,16, thus potentially observable by K2. 
  In Fig.~\ref{fig:star}, we present the light curve of the star EPIC 201872595
  (K$_{p}$ = 12.2) from Campaign \#1, in which each surge of flux is caused by 
  the transit of a different SSO within the target mask. The stellar flux is 
  clearly contaminated by the SSOs. This is an obvious case of transits by SSOs, 
  each being barely less bright (V\,$\sim$\,14\,--\,15) than the target star. 
  Fainter SSOs (V\,$\sim$\,18\,--\,19) still affect stellar light curves, without 
  being easily identifiable by naked eye. Using the SkyBoT service, it is easy to
  check any suspicious point in a stellar light curve, by performing a cone-search, 
  centered on the star, at the time of the corresponding photometry measurement, 
  with a narrow field of view of a few arcseconds corresponding to the apparent 
  size of the stellar mask.\\ 

  \indent The service also allows to hunt for photometric data of SSOs. One can 
  use SkyBoT to get the list of all the SSOs within the K2 entire FoV for each 
  campaign, and compute their encounters with target stars to extract their 
  photometry. For the fast generation of detailed ephemerides for each target, 
  we recommend the use of our Miriade service \citep[][]{2009epsc.conf..676B-Berthier}.
  Requesting SkyBoT cone-search for the entire FoV, with a time step of 30 min 
  during a whole campaign, is more CPU intensive than computing the same 
  ephemerides for only the identified targets with Miriade.\\

  \indent In Fig.~\ref{fig:sso} we present 10 light curves of asteroid
  (484) Pittsburghia (apparent magnitude $\sim$15) we measured in K2 
  Campaign \#0. The light curves have been constructed following the steps
  described above: a global SkyBoT request, followed by a Miriade generation 
  of ephemerides every 30 min for Pittsburghia, and finally a check of 
  whenever the asteroid was within one of the stellar masks. The synthetic 
  light curve was generated using the 3-D shape model of Pittsburghia by 
  \citet{2009-AA-493-Durech} and \citet{2011-AA-530-Hanus} is overplotted 
  to the data. The excellent match of the photometry measured on K2 frames 
  with the shape models illustrate the interest of data mining K2 data 
  archive for SSO period determination, and shape modeling.

\section{Conclusion}

  \indent We present a new version of the Virtual Observatory Web service 
  SkyBoT. Its cone-search method allows to list all the solar system objects 
  present within a given field of view at a given epoch, as visible from the 
  Earth, the ESA \textit{Rosetta} mission, and now the NASA \textit{Kepler} telescope. 
  More space missions can be added upon request, if desired by the community. 
  Typical queries over limited field of views take less than 10\,s, while 
  queries over extended field of view such as \textit{Rosetta}/OSIRIS camera 
  or \textit{Kepler} full CCD array take about a minute. Possible applications 
  of SkyBoT for K2 data are presented, and the results illustrate the interest 
  of K2 for studying asteroids spin, period, and shapes from the light curves
  which can be extracted from K2 data. Their analysis and interpretation 
  will be presented in a forthcoming paper (Carry et al., in preparation).

\section*{Acknowledgements}

  We acknowledges support of the ESAC Faculty for J.~Berthier's visit.
  This research has received funding from the European Union's H2020-
  PROTEC-2014 - Protection of European assets in and from space project 
  no. 640351 (NEOShield-2). A. Santerne is supported by the European Union 
  under a Marie Curie Intra-European Fellowship for Career Development with 
  reference FP7-PEOPLE-2013-IEF, number 627202. He also acknowledges the 
  support from the Funda\c{c}\~ao para a Ci\^encia e Tecnologia, FCT (Portugal) 
  in the form of the grants UID/FIS/04434/2013 (POCI-01-0145-FEDER-007672) 
  and POPH/FSE (EC) by FEDER funding through the program ``Programa Operacional 
  de Factores de Competitividade - COMPETE''.



\bibliographystyle{mnras}

\bsp	
\label{lastpage}
\end{document}